\UseRawInputEncoding

\documentclass[final,3p,times,sort&compress]{elsarticle}

\usepackage{epstopdf}
\usepackage{lineno}
\usepackage{amsmath}
\usepackage{bm}
\usepackage{amssymb}
\usepackage{setspace}
\usepackage[normalem]{ulem}
\usepackage{color}

\journal{Probabilistic Engineering Mechanics}

\begin{document}

\begin{frontmatter}


\title{Comparison of robust, reliability-based and non-probabilistic topology optimization under uncertain loads and stress constraints}
\author[cor1]{Gustavo Assis da Silva}
\ead{gustavoas@usp.br}
\author[cor2]{Eduardo Lenz Cardoso}
\ead{eduardo.cardoso@udesc.br}
\author[cor1]{Andr\'e Te\'ofilo Beck}
\ead{atbeck@sc.usp.br}
\address[cor1]{Department of Structural Engineering, S\~{a}o Carlos School of Engineering, University of S\~{a}o Paulo, 13.566-590, S\~{a}o Carlos, SP, Brazil}
\address[cor2]{Department of Mechanical Engineering, State University of Santa Catarina, 89.219-710, Joinville,
SC, Brazil}





\begin{abstract}

It is nowadays widely acknowledged that optimal structural design should be robust with respect to the uncertainties in loads and material parameters. However, there are several alternatives to consider such uncertainties in structural optimization problems. This paper presents a comprehensive comparison between the results of three different approaches to topology optimization under uncertain loading, considering stress constraints: 1) the robust formulation, which requires only the mean and standard deviation of stresses at each element; 2) the reliability-based formulation, which imposes a reliability constraint on computed stresses; 3) the non-probabilistic formulation, which considers a worst-case scenario for the stresses caused by uncertain loads. The information required by each method, regarding the uncertain loads, and the uncertainty propagation approach used in each case is quite different. The robust formulation requires only mean and standard deviation of uncertain loads; stresses are computed via a first-order perturbation approach. The reliability-based formulation requires full probability distributions of random loads, reliability constraints are computed via a first-order performance measure approach. The non-probabilistic formulation is applicable for bounded uncertain loads; only lower and upper bounds are used, and  worst-case stresses are computed via a nested optimization with anti-optimization. The three approaches are quite different in the handling of uncertainties; however, the basic topology optimization framework is the same: the traditional density approach is employed for material parameterization, while the augmented Lagrangian method is employed to solve the resulting problem, in order to handle the large number of stress constraints. Results are computed for two reference problems: similarities and differences between optimized topologies obtained with the three formulations are exploited and discussed.

\end{abstract}


\end{frontmatter}



\section{Introduction}

Handling uncertainties in structural optimization is fundamental in order to design high performance structures which are reliable and insensitive to parameters which may affect the structural response \cite{Melchers_Beck}. Neglecting the effect of such uncertainties during the structural optimization process may be catastrophic, leading to sub-optimal or even non-optimal designs, which may be unreliable and too sensitive with respect to the uncertain parameters \cite{Melchers_Beck}.

In order to overcome this issue and allow the proper handling of uncertainties in structural optimization, several works on structural optimization under uncertainties have been developed (see references \cite{Elishakoff_1994,Sorensen_1994,Lombardi,Lombardi_2,Jensen_2006,Aoues_2008,Guest_coordenadas,Guo_2009_global,Jensen_2011,Guest_trelica_2011,Beck2012,Beck2015,Bucher_2018}, for instance). The formulations addressed in these works are mostly based on established frameworks for optimization under uncertainty, such as: 1) probabilistic robust approach \cite{BeyerRDO}, where statistical moments are considered during optimization, as expectations and standard deviations, aiming at reducing sensitivity of the optimized design with respect to the uncertain variables; 2) probabilistic reliability-based approach \cite{RBDO_rev_2}, developed to ensure an admissible failure probability to the optimized design; 3) non-probabilistic robust approach \cite{Elishakoff_AOTM}, based on the worst-case scenario for the design constraints, employed when there is only the interval information for the unknown variables.

In this paper, a special category of structural optimization is addressed: topology optimization of solid structures. Topology optimization of continuum structures aims to find the best material distribution in a given fixed domain in order to extremize an objective function while respecting a given set of constraints. Topology optimization is the most general category of structural optimization, since it determines the optimal number, shape and position of holes \cite{BendsoeLivro}.

Among the several structural engineering problems which may be addressed through topology optimization, there is the classical minimum weight problem subjected to stress failure criterion \cite{DuysinxBendsoe}. Due to their extreme importance and applicability from an engineering point of view, stress-constrained topology optimization problems are subject of intensive research in the literature (see references \cite{DuysinxBendsoe,DuysinxSigmund,Fancello_Tensao,Pereira2004,Fancello_2006,Bruggi2008,Paris_SMO,ChauLe,Novotny_1,Novotny_2,Fancello_Level_set1,Fancello_Level_set2,IVE_Tensao,DeLeon2015,Alexandre_mecanismos,Norato_level_set,Artigo2,Artigo3,Artigo4,Artigo5,Artigo6,RBTO_Tensao,RTO_Game_Theory,Holmberg_RTO_2,novotny_rbto_tensao}, for instance). Among these works on stress-based design, there are some papers addressing the effect of uncertainties in applied loads, material properties and geometric parameters \cite{Artigo2,Artigo3,Artigo4,Artigo5,Artigo6,RBTO_Tensao,RTO_Game_Theory,Holmberg_RTO_2,novotny_rbto_tensao}.

In this paper, the minimum weight problem subjected to stress failure criterion is addressed. This paper aims at comparing three distinct formulations for stress-constrained topology optimization under load uncertainties: two probabilistic and one non-probabilistic approach. Each formulation is classified depending on the uncertainty quantification and propagation approaches employed for handling the uncertainties:
\begin{enumerate}
\item Probabilistic robust approach, proposed in \cite{Artigo3}: only mean and (co)variance of random loads is considered; stress constraints are written as a weighted sum between expectation and standard deviation;
\item Probabilistic reliability-based approach, proposed in \cite{Artigo4}: full probabilistic information is used; stress constraints are written in terms of acceptable probabilities of occurrence;
\item Non-probabilistic robust approach, proposed in \cite{Artigo5}: for bounded uncertainty in loads; worst-case scenario for stress constraints are considered.
\end{enumerate}

For the sake of simplicity, approaches 1, 2 and 3 are called as robust, reliability-based and non-probabilistic, throughout the paper, respectively, in order to avoid misunderstandings.

The remainder of this paper is organized as follows: the stress-constrained formulations (both deterministic and non-deterministic) are presented in section \ref{formulacoes}; the method employed to solve the optimization problems is presented in section \ref{procedimento_solucao}; optimization examples and discussions are shown in section \ref{resultados}; and concluding remarks are given in section \ref{conclusoes}. Additional insight about the density-based framework for topology optimization is given in the appendix.

\section{Stress-constrained topology optimization}
\label{formulacoes}

In this paper, the traditional density approach \cite{Sigmund2013} is employed as topology optimization framework: 1) the continuum design domain is discretized with finite elements \cite{Bathe}; 2) each finite element $e$ is associated with a relative density $\overline{\rho}_{e} \in [0,1]$, where $0$ represents void and $1$ represents solid material. After defining the structural optimization problem (e.g., minimum compliance design \cite{Bendsoe1999}, stress-based design \cite{ChauLe}, compliant mechanism design \cite{Sigmund_1997}), established optimization methods (see references \cite{Rao2009,Arora2012,MMA}, for instance) are employed to update the relative densities and find an optimized configuration which minimizes (or maximizes) the objective function, respecting the applied design constraints.

Next subsections are devoted to present and explain the formulations addressed in this paper, based on the volume minimization problem subjected to local stress constraints. Subsection \ref{DTO} presents the deterministic formulation, whereas subsections \ref{RTO}, \ref{RBTO} and \ref{nRTO} present the non-deterministic formulations, developed to address the same structural problem under the effect of uncertainties in applied loads. All problems are formulated by employing the same base interpolation functions and governing parameters (e.g., stiffness, volume and stress interpolation functions), as well as behavior hypotheses, as defined for the deterministic problem in subsection \ref{DTO}. Moreover, each non-deterministic approach has its own particularities, which are beyond the deterministic formulation. These particularities are presented and explained in each respective subsection.

\subsection{Deterministic approach}
\label{DTO}

Considering that the equilibrium configuration of the structural problem is obtained with the displacement-based finite element method for linear elasticity under static loads \cite{Bathe}, and adopting the classical von Mises stress failure criterion, one can write the deterministic problem, in discrete form, as
\begin{linenomath}\begin{equation}
\begin{array}{lll}
\begin{array}{cc} \vspace{-12pt} \overset{\displaystyle \mathrm{Min.}}{^{\bm{\rho}}} \end{array} & V_{p}(\overline{\bm{\rho}}) = \sum_{e=1}^{N_{e}} V_{e} f_{v}\left( \overline{\rho}_{e} \right) & \\ \\
\begin{array}{c} $ s. t.$ \end{array} & \frac{\sigma_{eq}^{(k)}(\overline{\bm{\rho}})}{\sigma_{y}} - 1 \leqslant 0 & \quad k=1,2,...,N_{k} \\
 & \mathbf{K}(\overline{\bm{\rho}})\mathbf{U}(\overline{\bm{\rho}}) = \mathbf{F} & \\ 
 & 0 \leqslant \rho_{e} \leqslant 1 & \quad e=1,2,...,N_{e}
\end{array},\label{Problema_otm_1}
\end{equation}\end{linenomath}
where $\bm{\rho} \in \mathbb{R}^{N_{e}}$ are the design variables of the optimization problem, $V_{p}(\overline{\bm{\rho}})$ is the objective function of the optimization problem (a penalized version of the structural volume $V(\overline{\bm{\rho}}) = \sum_{e=1}^{N_{e}} V_{e} \overline{\rho}_{e}$), which depends on the physical relative densities $\overline{\bm{\rho}} \in \mathbb{R}^{N_{e}}$, $N_{e}$ is the number of finite elements comprising the design domain, $V_{e}$ is the structural volume of finite element $e$, $f_{v}\left( \overline{\rho}_{e} \right)$ is the volume penalization function evaluated at element $e$, $\sigma_{eq}^{(k)}(\overline{\bm{\rho}})$ is the von Mises equivalent stress at point $k$, $\sigma_{y}$ is the yield stress of solid material, $N_{k}$ is the number of points where the von Mises equivalent stress is computed, $\mathbf{K}(\overline{\bm{\rho}})$ is the global stiffness matrix, $\mathbf{U}(\overline{\bm{\rho}})$ is the global displacement vector and $\mathbf{F}$ is the global load vector. The local stiffness matrix of element $e$ is interpolated by adopting the Solid Isotropic Material with Penalization (SIMP) approach, as $\mathbf{k}_{e}(\overline{\rho}_{e}) = \left( \overline{\rho}_{e}^{p} + \rho_{min} \right) \mathbf{k}_{e}^{b}$, following \cite{Guest_beta_fixo} and \cite{Guest_Spectral}, where $\rho_{min} = 1 \times 10^{-9}$ is adopted to ensure a well-conditioned system of linear equations, $p>1$ is a penalization factor (often chosen as $p=3$ in the literature, and also in this paper), and $\mathbf{k}_{e}^{b}$ is the stiffness matrix considering solid material.

The volume penalization function is chosen as $f_{v}\left( \overline{\rho}_{e} \right) = 1 - \mathrm{e}^{-\delta_{v} \overline{\rho}_{e}} + \overline{\rho}_{e} \mathrm{e}^{-\delta_{v}}$, with $\delta_{v} = 5$, in order to penalize relative densities and make intermediate material uneconomical to the optimizer, following \cite{Artigo4,Artigo5}.

The von Mises equivalent stress at any point $k$ is computed based on \cite{DuysinxBendsoe}, and can be written as
\begin{linenomath}\begin{equation}
\sigma_{eq}^{(k)}\left(\overline{\bm{\rho}}\right) = \sqrt{\bm{\sigma}_{k}^{T}\left(\overline{\bm{\rho}}\right) \mathbf{M} \bm{\sigma}_{k}\left(\overline{\bm{\rho}}\right) + \sigma_{min}^2},\label{tensao_von_Mises_2}
\end{equation}\end{linenomath}
where the constant $\sigma_{min} = 1 \times 10^{-4} \sigma_{y}$ is included in our implementations to ensure a positive von Mises equivalent stress when $\bm{\sigma}_{k}^{T}\left(\overline{\bm{\rho}}\right) \mathbf{M} \bm{\sigma}_{k}\left(\overline{\bm{\rho}}\right) \rightarrow 0$, in order to avoid numerical instabilities during the sensitivity analysis, needed for optimization with gradient-based algorithm.

In Equation \eqref{tensao_von_Mises_2}, the stress vector at point $k$, $\bm{\sigma}_{k}\left(\overline{\bm{\rho}}\right)$, is computed as
\begin{linenomath}\begin{equation}
\bm{\sigma}_{k}\left(\overline{\bm{\rho}}\right) = \overline{\mathbf{C}}(\overline{\rho}_{k}) \mathbf{B}_{k} \mathbf{u}_{k}(\overline{\bm{\rho}}),\label{tensao}
\end{equation}\end{linenomath}
and the matrix $\mathbf{M}$, for plane stress problems, is defined as
\begin{linenomath}\begin{equation}
\mathbf{M} = \left[\begin{array}{ccc}
1 & -0.5 & 0 \\ 
-0.5 & 1 & 0 \\ 
0 & 0 & 3
\end{array} \right].\label{matriz_M}
\end{equation}\end{linenomath}

In Equation \eqref{tensao}, $\overline{\mathbf{C}}(\overline{\rho}_{k}) = f_{\sigma} \left( \overline{\rho}_{k} \right) \mathbf{C}^{b}$ is a modified constitutive matrix, which is computed by interpolating the constitutive matrix of solid material, $\mathbf{C}^{b}$, by a stress interpolation function, $f_{\sigma} \left( \overline{\rho}_{k} \right)$; $\mathbf{B}_{k}$ is the strain-displacement transformation matrix evaluated at point $k$; and $\mathbf{u}_{k}(\overline{\bm{\rho}})$ is the local displacement vector of the element which contains point $k$.

In this work we choose $f_{\sigma} \left( \overline{\rho}_{k} \right) = 1 - \mathrm{e}^{-\delta_{\sigma} \overline{\rho}_{k}} + \overline{\rho}_{k} \mathrm{e}^{-\delta_{\sigma}}$ to interpolate the stresses, in order to relax the stress constraints and ensure automatic stress constraint feasibility for $\overline{\rho}_{k} \rightarrow 0$, thus avoiding the singularity phenomenon \cite{DuysinxBendsoe,ChengGuo}. This choice holds for deterministic and non-deterministic formulations, with $\delta_{\sigma} = 3$, following \citep{Artigo4,Artigo5}.

Relative densities $\overline{\bm{\rho}}$ are related to design variables $\bm{\rho}$ through density filtering with Heaviside step function \citep{Sigmund-2007}, as shown in the appendix.

\subsection{Robust approach}
\label{RTO}

Considering the design domain is under the effect of random external loads, one has to reformulate the originally deterministic optimization problem, Equation \eqref{Problema_otm_1}, in order to properly take these uncertainties into account. The robust formulation addressed in this work was proposed in \cite{Artigo3}, and consists in the replacement of each original deterministic stress constraint by a weighted sum between its expectation and standard deviation. The robust formulation, in discrete form, is written as
\begin{linenomath}\begin{equation}
\begin{array}{lll}
\begin{array}{cc} \vspace{-12pt} \overset{\displaystyle \mathrm{Min.}}{^{\bm{\rho}}} \end{array} & V_{p}(\overline{\bm{\rho}}) & \\ \\
\begin{array}{c} $ s. t.$ \end{array} & \frac{\hat{\sigma}_{eq}^{(k)}(\overline{\bm{\rho}},\mathbf{Z})}{\sigma_{y}} - 1 \leqslant 0 & \quad k=1,2,...,N_{k} \\
 & \mathbf{K}(\overline{\bm{\rho}})\mathbf{U}(\overline{\bm{\rho}},\mathbf{Z}) = \mathbf{F}(\mathbf{Z}) & \\ 
 & 0 \leqslant \rho_{e} \leqslant 1 & \quad e=1,2,...,N_{e}
\end{array},\label{Problema_otm_2}
\end{equation}\end{linenomath}
where $\mathbf{Z} \in \mathbb{R}^{N}$ is a vector containing all random loads of the problem. The stress measure, based on the von Mises equivalent stress, which is considered during the optimization process, is computed as
\begin{equation}
\hat{\sigma}_{eq}^{(k)}(\overline{\bm{\rho}},\mathbf{Z}) = \mathrm{E} \left[ \sigma_{eq}^{(k)}(\overline{\bm{\rho}},\mathbf{Z}) \right] + \alpha \; \mathrm{Std} \left[ \sigma_{eq}^{(k)}(\overline{\bm{\rho}},\mathbf{Z}) \right],\label{tensao_RTO}
\end{equation}
where $\mathrm{E} [\cdot]$ represents the expected value and $\mathrm{Std} [\cdot]$ the standard deviation of the von Mises equivalent stresses, which depend on the random loads $\mathbf{Z}$, and $\alpha$ is a weighting parameter, which should be adjusted by the designer in order to ensure a desired degree of robustness to the optimized topology.

Parameter $\alpha$, in Equation \eqref{tensao_RTO}, should not be confused with the target reliability index, $\beta_{T}$, often adopted in reliability-based formulations based on first-order approximations. However, as discussed in \cite{Artigo2}, the value of $\alpha$ can be properly chosen to ensure a conservative limit of the probability of failure by employing the one-sided Chebyshev inequality.

The expectation $\mathrm{E} [\cdot]$ and standard deviation $\mathrm{Std} [\cdot]$ of von Mises equivalent stresses may be computed by employing any uncertainty propagation technique usually found in the literature, as the Monte Carlo Simulation (MCS) \cite{Melchers_Beck}. However, since we are handling with extremely challenging optimization problems, with thousands of design variables and thousands of stress constraints (as presented later in the results section), we preferred to employ a cheaper first-order perturbation approach for uncertainty quantification. Following \cite{Artigo3}, expectation and variance of von Mises equivalent stress, at point $k$, are computed as
\begin{equation}
\mathrm{E}\left[\sigma_{eq}^{(k)}(\overline{\bm{\rho}},\mathbf{Z})\right] \cong \left.\sigma_{eq}^{(k)}\left(\overline{\bm{\rho}},\mathbf{Z}\right)\right|_{\mathbf{Z}=\mathrm{E}[\mathbf{Z}]},\label{valor_esperado}
\end{equation}
and
\begin{equation}
\mathrm{Var}\left[\sigma_{eq}^{(k)}(\overline{\bm{\rho}},\mathbf{Z})\right] \cong \sum_{i=1}^{N} \sum_{j=1}^{N} \left.\frac{\partial \sigma_{eq}^{(k)}}{\partial Z_{i}}\right|_{\mathbf{Z}=\mathrm{E}[\mathbf{Z}]} \left.\frac{\partial \sigma_{eq}^{(k)}}{\partial Z_{j}}\right|_{\mathbf{Z}=\mathrm{E}[\mathbf{Z}]} \mathrm{Cov}(Z_{i},Z_{j}),\label{variancia}
\end{equation}
respectively, where $\mathrm{Cov}(Z_{i},Z_{j})$ represents the covariance between random variables $Z_{i}$ and $Z_{j}$. After computing the variance, the standard deviation of von Mises stress is simply computed as
\begin{equation}
\mathrm{Std}\left[\sigma_{eq}^{(k)}\left(\overline{\bm{\rho}},\mathbf{Z}\right)\right] = \sqrt{\mathrm{Var}\left[\sigma_{eq}^{(k)}\left(\overline{\bm{\rho}},\mathbf{Z}\right)\right] + \sigma_{min}^2},\label{desvio_padrao}
\end{equation}
where the small number $\sigma_{min}^2$ is included in our implementations to avoid numerical instabilities during the sensitivity analysis.

The authors refer the reader to \cite{Artigo3} for further insight about the robust formulation, as well as the development of necessary derivatives for analytical evaluation of expectation and variance of von Mises equivalent stresses, Equations \eqref{valor_esperado} and \eqref{variancia}, respectively. The robust solution above only requires mean, $\mathrm{E}[\mathbf{Z}]$, and covariance, $\mathrm{Cov}(Z_{i},Z_{j})$, of random problem parameters.

\subsection{Reliability-based approach}
\label{RBTO}

The reliability-based formulation addressed in this work is proposed in \cite{Artigo4}, in which probability of occurrence of von Mises stress, at each point of stress evaluation, is constrained by a given admissible failure probability, $P_{adm}$. The reliability-based formulation, in discrete form, is written as
\begin{linenomath}\begin{equation}
\begin{array}{lll}
\begin{array}{cc} \vspace{-12pt} \overset{\displaystyle \mathrm{Min.}}{^{\bm{\rho}}} \end{array} & V_{p}(\overline{\bm{\rho}}) & \\ \\
\begin{array}{c} $ s. t.$ \end{array} & P \left( \frac{\sigma_{eq}^{(k)}(\overline{\bm{\rho}},\mathbf{Z})}{\sigma_{y}} - 1 \geqslant 0 \right) \leqslant P_{adm} & \quad k=1,2,...,N_{k} \\
 & \mathbf{K}(\overline{\bm{\rho}})\mathbf{U}(\overline{\bm{\rho}},\mathbf{Z}) = \mathbf{F}(\mathbf{Z}) & \\ 
 & 0 \leqslant \rho_{e} \leqslant 1 & \quad e=1,2,...,N_{e}
\end{array},\label{Problema_otm_3}
\end{equation}\end{linenomath}
where $P(\cdot)$ represents probability.

In this paper, the reliability-based optimization problem, as defined in Equation \eqref{Problema_otm_3}, is not directly solved, due to the impracticability of analytical evaluation of such probabilities. Among the several existing techniques proposed in the literature for numerically addressing reliability-based problems \cite{RBDO_rev_1,RBDO_rev_2,LopezBeckRBDO}, we choose to employ the Performance Measure Approach (PMA) \cite{Tu_PMA}, following \cite{Artigo4}.

The PMA consists in a nested strategy: at each time the topology is updated (iteration of outer optimization problem), the Minimal Performance Points (MiPPs) must be found by performing one first-order inverse reliability analysis in standard normal space (inner optimization problem), for each stress constraint.

The outer optimization problem is written as
\begin{linenomath}\begin{equation}
\begin{array}{lll}
\begin{array}{cc} \vspace{-12pt} \overset{\displaystyle \mathrm{Min.}}{^{\bm{\rho}}} \end{array} & V_{p}(\overline{\bm{\rho}}) & \\ \\
\begin{array}{c} $ s. t.$ \end{array} & \frac{\sigma_{eq}^{(k)}\left(\overline{\bm{\rho}},\left(\mathbf{Y}^{(k)}\right)^{*}\right)}{\sigma_{y}} - 1 \leqslant 0 & \quad k=1,2,...,N_{k} \\
 & \mathbf{K}(\overline{\bm{\rho}})\mathbf{U}(\overline{\bm{\rho}},\mathbf{Y}) = \mathbf{F}(\mathbf{Y}) & \\ 
 & 0 \leqslant \rho_{e} \leqslant 1 & \quad e=1,2,...,N_{e}
\end{array},\label{Problema_otm_4}
\end{equation}\end{linenomath}
where $\mathbf{Y} \in \mathbb{R}^{N}$ is a vector containing all random variables of the problem in standard normal space $\mathbb{Y}$, and $\left(\mathbf{Y}^{(k)}\right)^{*}$ is the MiPP associated with $k$-th stress constraint.

The inner optimization problem (first-order inverse reliability analysis) is defined, for each stress constraint, as the minimization of the negative of the constraint function by adopting the random variables $\mathbf{Y}$ as design variables. The $k$-th inner problem is written as
\begin{linenomath}\begin{equation}
\begin{array}{lll}
\begin{array}{cc} \vspace{-12pt} \overset{\displaystyle \mathrm{Min.}}{^{\mathbf{Y}}} \end{array} & -\sigma_{eq}^{(k)}\left(\overline{\bm{\rho}},\mathbf{Y}\right) \\ \\
\begin{array}{c} $ s. t.$ \end{array} & \Vert \mathbf{Y} \Vert = \beta_{T}
\end{array},\label{Problema_otm_5}
\end{equation}\end{linenomath}
where $\Vert \mathbf{Y} \Vert$ is the Euclidean norm of $\mathbf{Y}$, which defines the size of the target reliability index
hyper-sphere. In order to evaluate $\sigma_{eq}^{(k)}\left(\overline{\bm{\rho}},\left(\mathbf{Y}^{(k)}\right)^{*}\right)$, in Equation \eqref{Problema_otm_4}, for $k=1,2,...,N_{k}$, one has to solve $N_{k}$ inner optimization problems, Equation \eqref{Problema_otm_5} (one for each design constraint). In this paper, inner optimization problems are solved with the Hybrid Mean Value (HMV) algorithm, as originally proposed by \citep{Youn_HMV}.

In the PMA, the target reliability index, $\beta_{T}$, is related to admissible failure probability through the standard Gaussian cumulative distribution function $\Phi$, such that $\beta_{T} \cong - \Phi^{-1} \left( P_{adm} \right)$, which consists in an approximate relation consistent with first-order approaches \citep{LopezBeckRBDO}.

It should be noted that inner problems are solved in standard normal space; thus, a transformation must be performed from original space $\mathbb{Z}$, which can be non Gaussian, to the standard normal space $\mathbb{Y}$. There are some techniques which may be employed to perform this transformation, mapping from $\mathbb{Z}$ to $\mathbb{Y}$, and the reader may consult \cite{Ditlevsen} for details. In this work, all random variables are Gaussian, such that Hasofer and Lind transformation is sufficient \citep{LopezBeckRBDO}. Note that this solution requires full probability distribution information about random problem parameters.

The authors refer the reader to \cite{Artigo4} for further insight about the adopted reliability-based formulation, as well as development of fast solution based on the principle of superposition and necessary derivatives for solving the inner problems by employing the HMV algorithm.

\subsection{Non-probabilistic approach}
\label{nRTO}

In robust and reliability-based formulations, presented earlier in subsections \ref{RTO} and \ref{RBTO}, respectively, uncertainties in applied loads are described as random vectors $\mathbf{Z}$. Each component $Z_{i}$, of $\mathbf{Z}$, represents an uncertain magnitude or direction of an applied load, and may assume any probability distribution function. As an alternative to the probabilistic representation, a non-probabilistic representation is addressed herein, where the uncertainties in applied loads are described by unknown-but-bounded variables $\mathbf{W} \in [\underline{\mathbf{W}},\overline{\mathbf{W}}]$, i.e., in this case, the unknown variables which describe the uncertainties in magnitudes and/or directions of applied loads may assume any value between the prescribed lower and upper limits. The non-probabilistic formulation addressed in this work is proposed in \cite{Artigo5}, and is based on the worst-case scenario for the stress constraints.

The non-probabilistic approach employed herein is remarkably similar to the PMA, Equations \eqref{Problema_otm_4} and \eqref{Problema_otm_5}, in the sense that it also consists in a nested optimization loop. The non-probabilistic approach employed herein is based on the two-level optimization with anti-optimization approach, described in \cite{Elishakoff_AOTM}.

The outer optimization problem is written as
\begin{linenomath}\begin{equation}
\begin{array}{lll}
\begin{array}{cc} \vspace{-12pt} \overset{\displaystyle \mathrm{Min.}}{^{\bm{\rho}}} \end{array} & V_{p}(\overline{\bm{\rho}}) & \\ \\
\begin{array}{c} $ s. t.$ \end{array} & \frac{\sigma_{eq}^{(k)}\left(\overline{\bm{\rho}},\left(\mathbf{W}^{(k)}\right)^{*}\right)}{\sigma_{y}} - 1 \leqslant 0 & \quad k=1,2,...,N_{k} \\
 & \mathbf{K}(\overline{\bm{\rho}})\mathbf{U}(\overline{\bm{\rho}},\mathbf{W}) = \mathbf{F}(\mathbf{W}) & \\ 
 & 0 \leqslant \rho_{e} \leqslant 1 & \quad e=1,2,...,N_{e}
\end{array},\label{Problema_otm_6}
\end{equation}\end{linenomath}
where $\left(\mathbf{W}^{(k)}\right)^{*}$ is the optimum set of applied loads, which gives the maximum value for the von Mises equivalent stress at point $k$.

In the same way as presented earlier for the PMA (reliability-based problem), one optimum set of applied loads $\left(\mathbf{W}^{(k)}\right)^{*}$ should be obtained for each stress constraint, in order to obtain the worst-case scenario for the stress constraints at a given iteration of the outer optimization problem.

The $k$-th inner optimization problem (also called anti-optimization problem) is written as
\begin{linenomath}\begin{equation}
\begin{array}{lll}
\begin{array}{cc} \vspace{-12pt} \overset{\displaystyle \mathrm{Max.}}{^{\mathbf{W}}} \end{array} & \sigma_{eq}^{(k)}\left(\overline{\bm{\rho}},\mathbf{W}\right) \\ \\
\begin{array}{c} $ s. t.$ \end{array} & \underline{\mathbf{W}} \leqslant \mathbf{W} \leqslant \overline{\mathbf{W}}
\end{array}.\label{Problema_otm_7}
\end{equation}\end{linenomath}

Solutions of anti-optimization problems should be global minima in order to ensure a truly worst-case scenario for the stress constraints and guarantee stress feasibility for the optimized structure under any load condition given the prescribed bounds for the uncertain variables \cite{Guo_2009_global}. However, ensuring global optima in non-convex optimization is always a challenging task \cite{Arora2012}. In this paper, we employ a two step approach for solving the anti-optimization problems, following \cite{Artigo5}: a grid search method \cite{Rao2009} followed by a modified steepest descent method \cite{Artigo3}. Although there is no mathematical guarantee that the obtained solutions are global minima, this simple two step formulation demonstrated to provide good results \cite{Artigo5}.

The authors refer the reader to \cite{Artigo5} for further insight about the adopted non-probabilistic formulation, as well as the development of fast solution based on the principle of superposition and necessary derivatives for solving the inner problems by employing the proposed two step optimization approach.

\section{Solution procedure}
\label{procedimento_solucao}

The optimization problems in Equations \eqref{Problema_otm_1}, \eqref{Problema_otm_2}, \eqref{Problema_otm_4} and \eqref{Problema_otm_6}, which define the deterministic, robust, reliability-based and non-probabilistic topology optimization problems in discrete form, respectively, are solved by employing the augmented Lagrangian method \cite{Martinez}, following implementation described in \cite{Artigo3}. The inner problems, related to the reliability-based and non-probabilistic formulations, are solved with specific algorithms described in subsections \ref{RBTO} and \ref{nRTO}, respectively.

The augmented Lagrangian method consists in a sequential formulation: the original constrained optimization problem is replaced by a sequence of optimization subproblems. The objective function of the optimization subproblems is the augmented Lagrangian function, which consists in the original objective function (the penalized structural volume) weighted by the design constraints (von Mises stress constraints) and respective Lagrange multipliers. After solving a given optimization subproblem, the penalization parameter and Lagrangian multipliers are updated; then, the next optimization subproblem is solved; and so on, until convergence.

By analyzing Equations \eqref{Problema_otm_1}, \eqref{Problema_otm_2}, \eqref{Problema_otm_4} and \eqref{Problema_otm_6}, one can verify that the only difference among them is the evaluation of the stress measure which is actually employed during the optimization process: $\sigma_{eq}^{(k)}(\overline{\bm{\rho}})$ (deterministic); $\hat{\sigma}_{eq}^{(k)}(\overline{\bm{\rho}},\mathbf{Z})$ (robust); $\sigma_{eq}^{(k)}\left(\overline{\bm{\rho}},\left(\mathbf{Y}^{(k)}\right)^{*}\right)$ (reliability-based); and $\sigma_{eq}^{(k)}\left(\overline{\bm{\rho}},\left(\mathbf{W}^{(k)}\right)^{*}\right)$ (non-probabilistic). Thus, in this section, we show the augmented Lagrangian function for $\sigma_{eq}^{(k)}(\overline{\bm{\rho}})$, only, but it can be defined for any other stress measure in the same way, by replacing the employed stress measure.

The augmented Lagrangian function is defined considering all stress constraints of the optimization problem, such that:
\begin{align}
L\left(\overline{\bm{\rho}},\bm{\mu},r\right) = & \frac{N_{e}}{\sum_{e=1}^{N_{e}}V_{e}} V_{p}\left(\overline{\bm{\rho}}\right) + \frac{r}{2}\sum_{k=1}^{N_{k}} \left\langle \frac{\mu_{k}}{r} + \frac{\sigma_{eq}^{(k)}\left(\overline{\bm{\rho}}\right)}{\sigma_{y}} - 1 \right\rangle^{2},\label{LA}
\end{align}
where $\bm{\mu} \in \mathbb{R}^{N_{k}}$ is a vector which contains all Lagrange multipliers of the problem, $r$ is the penalization parameter, $\mu_{k}$ is the Lagrange multiplier associated with $k$-th stress constraint, and $\langle \cdot \rangle = \max(0,\cdot)$. The objective function is weighted by constant $\frac{N_{e}}{\sum_{e=1}^{N_{e}}V_{e}}$ for the purpose of normalization.

Since bound constraints are not included in the augmented Lagrangian function, they must be explicitly considered in the optimization subproblems, defined as
\begin{equation}
\begin{array}{lll}
\begin{array}{cc} \vspace{-12pt} \overset{\displaystyle \mathrm{Min.}}{^{\bm{\rho}}} \end{array} & L\left(\overline{\bm{\rho}},\bm{\mu}^{(c)},r^{(c)}\right) & \\  \\
\begin{array}{c} $ s. t.$ \end{array} & \mathbf{K}(\overline{\bm{\rho}})\mathbf{U}(\overline{\bm{\rho}}) = \mathbf{F} & \\ 
 & 0 \leqslant \rho_{e} \leqslant 1 & \quad e=1,2,...,N_{e}
\end{array},\label{Problema_otm_8}
\end{equation}
where the superscript $^{(c)}$ indicates $c$-th optimization subproblem.

After solving $c$-th optimization subproblem, one can employ the solution $\left(\bm{\rho}^{(c)}\right)^{*}$ of the current subproblem and current estimate of Lagrange multipliers $\bm{\mu}^{(c)}$ and penalization parameter $r^{(c)}$ to update the next estimate of Lagrange multipliers
\begin{linenomath}\begin{equation}
\mu_{k}^{(c+1)} \leftarrow \left\langle r^{(c)} \left(\frac{\sigma_{eq}^{(k)}\left(\left(\overline{\bm{\rho}}^{(c)}\right)^{*}\right)}{\sigma_{y}} - 1\right) + \mu_{k}^{(c)} \right\rangle,\label{Multiplicador_de_Lagrange}
\end{equation}\end{linenomath}
and penalization parameter
\begin{linenomath}\begin{equation}
r^{(c+1)} \leftarrow \left\lbrace \begin{array}{lll}
\min\left(\gamma \ r^{(c)},r_{max}\right) &  & $if $ \delta \sigma_{max}^{(c)} > \omega \ \delta \sigma_{max}^{(c-1)} \\
r^{(c)} &  & $otherwise$
\end{array}  \right.,\label{Parametro_de_penalizacao}
\end{equation}\end{linenomath}
where $\gamma>1$ and $\omega < 1$ are update parameters, $r_{max}$ is an upper value for the penalization parameter and $\delta \sigma_{max} = \left(\frac{\sigma_{eq}^{max}}{\sigma_{y}} - 1\right)$, where $\sigma_{eq}^{max}$ represents the maximum value among all computed von Mises equivalent stresses. The value of penalization parameter $r$ is increased by a factor of $\gamma$ only if the maximum value of stress constraints does not reduce at least by a factor of $\omega$, i.e., if there is reasonable progress regarding feasibility of the optimized topology, the penalization parameter is not updated, in order to avoid unnecessary increase of optimization subproblems nonlinearity.

This procedure is performed until both convergence criteria are simultaneously reached: 1) when maximum change on design variables becomes smaller than $tol_{out}$; 2) when feasibility is guaranteed, such that $\frac{\sigma_{eq}^{max}}{\sigma_{y}} - 1 < tol_{\sigma}$.

The value of $\delta$, from the smoothed Heaviside function, Equation \eqref{projecao_Heaviside}, employed after filtering design variables (see the appendix for details), is increased through a continuation approach: the problem is initially solved considering $\delta = 0$; then, the value of $\delta$ is increased as $\delta \leftarrow \delta + 5$ and next problem is solved, considering solution of current problem, current Lagrange multipliers and current penalization parameter as initial estimates, and so on, until an upper value $\delta_{max} = 100$ is reached. Large value of $\delta$ is employed to reduce blurred boundaries effect and achieve crisp black and white topologies.

The optimization subproblems, Equation \eqref{Problema_otm_8}, are solved with a modified steepest descent algorithm, described in \cite{Artigo3}. The necessary derivatives are developed by employing the adjoint technique and are presented in the base papers addressed in this work: 1) sensitivity analyses for deterministic and robust approaches \cite{Artigo3}; 2) sensitivity analysis for reliability-based approach \cite{Artigo4}; 3) sensitivity analysis for non-probabilistic approach \cite{Artigo5}.

\section{Numerical results and discussions}
\label{resultados}

\begin{figure*}[ht!]
\centering
\includegraphics[width=0.9\textwidth]{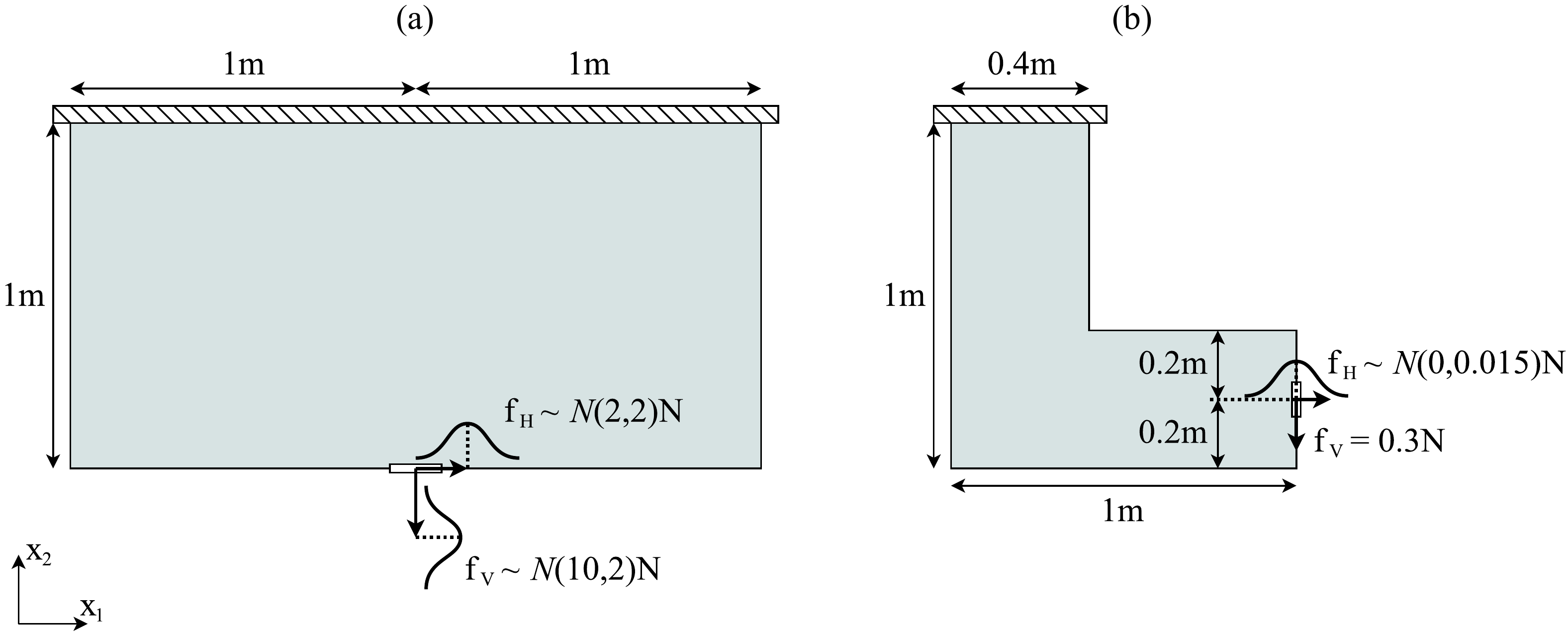}
\caption{Design domains with geometric dimensions and boundary conditions.}
\label{dominio_retangular_L1}
\end{figure*}

The numerical examples addressed in this section demonstrate the main similarities and differences among the three non-deterministic approaches presented in the paper. The numerical examples consist in 2D topology optimization problems, where hypotheses of plane stress are considered. Two problems are addressed: (a) problem with rectangular design domain under two uncertain loads; (b) L-shaped design problem under one deterministic and one uncertain load; Figure \ref{dominio_retangular_L1}. Obtained results are compared with the deterministic results, which are obtained for applied loads evaluated at their mean values.

Material and geometric parameters shared by both problems: Young's modulus of $1$MPa, thickness of $1$mm, and Poisson's ratio of $0.3$.

Input data for the optimization solver are: $r^{(1)} = 0.01$, $r_{max} = 10000$, $\gamma = 10$ and $\omega = 0.8$, as input data for the augmented Lagrangian method; $tol_{out} = 0.1$ and $tol_{\sigma} = 0.01$ as maximum change on design variables and required feasibility for the stress constraints, respectively; and $\bm{\rho} = \mathbf{1}$ as initial estimate for the design variables. The optimization subproblems are solved with a modified steepest descent method, proposed in \cite{Artigo3}, and it employs: maximum range of moving limits of $\rho_{e} \pm 0.1$; minimum range of moving limits of $\rho_{e} \pm 0.02$; tolerance based on maximum change on design variables of $tol_{sub} = 0.01$; and maximum number of iterations for a given subproblem of $nit_{max}=50$. Moving limits are heuristically updated based on two previous iterations, where parameters $k_{1} = 0.7$ and $k_{2} = 1.1$ are used for reducing moving limits if oscillation of design variables occur and increasing otherwise, respectively \citep{Artigo3}. At the beginning of each subproblem, moving limits are set to either maximum range ($\pm 0.1$), if $\delta = 0$, or minimum range ($\pm 0.02$), if $\delta > 0$, where $\delta$ governs the nonlinearity of the smoothed Heaviside projection, Equation \eqref{projecao_Heaviside}.

\begin{figure}[ht!]
\centering
\includegraphics[width=0.3\textwidth]{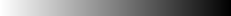}
\caption{Gray scale employed to represent topologies. White represents the void phase ($\overline{\rho} = 0$) and black the solid phase ($\overline{\rho} = 1$).}
\label{cores_densidades}
\end{figure}

\begin{figure}[ht!]
\centering
\includegraphics[width=0.3\textwidth]{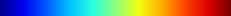}
\caption{Color scale employed to represent normalized von Mises stresses and post-processed reliability indices. In stress graphs: red represents maximum normalized stress ($\cong 1$) and blue the minimum normalized stress ($\cong 0$). In reliability indices graphs: red represents the minimum ($\beta_{min}$) and blue the maximum ($\beta_{max}$) post-processed reliability index.}
\label{cores}
\end{figure}

Additional data: the problems are discretized with square bilinear isoparametric finite elements, and the stresses are computed at the centroid of each element. Topologies are illustrated in gray scale, Figure \ref{cores_densidades}, where black represents solid material ($\overline{\rho} = 1$) and white represents void ($\overline{\rho} = 0$). Von Mises stresses are illustrated in color images, Figure \ref{cores}, where red represents maximum normalized stresses ($\cong 1$) and blue represents minimum normalized stresses ($\cong 0$). Post-processed reliability indices are illustrated in color images, Figure \ref{cores}, where red represents minimum reliability indices, $\beta_{min}$, and blue represents maximum reliability indices, $\beta_{max}$.

\subsection{Rectangular problem under two uncertain loads}

The problem with rectangular design domain, Figure \ref{dominio_retangular_L1} (a), is discretized with a finite element mesh of $80000$ elements. Input data: filtering radius of $R=0.04$m and yield stress of $\sigma_{y} = 100$kPa. Two applied loads of uncertain magnitude and deterministic direction are considered, one horizontal and one vertical. Both loads have Gaussian magnitude, $f \sim N \left( \text{E}[f] , \text{Std}[f] \right)$. The expected value of the horizontal load is $2$N, and its standard deviation is $2$N, such that $f_{H} \sim N (2,2)$N. The expected value of the vertical load is $10$N, and its standard deviation is $2$N, such that $f_{V} \sim N (10,2)$N. The random variables $f_{H}$ and $f_{V}$ are uncorrelated. Applied loads are distributed over a length of $0.2$m to avoid stress concentration.

Obviously, since uncertain loads with Gaussian magnitudes are considered, the most proper approaches that may be employed to formulate the optimization problems are the probabilistic ones. However, aiming at comparing the deterministic and the three non-deterministic (probabilistic and non-probabilistic) approaches, three different situations are considered, based on how the applied loads are handled:
\begin{itemize}
\item Deterministic problem, with $f_{V} = 10$N and $f_{H} = 2$N (applied loads are evaluated at the mean values);
\item Probabilistic problem, where applied loads are uncorrelated Gaussian variables as defined earlier, i.e., $f_{V} \sim N (10,2)$N and $f_{H} \sim N (2,2)$N. Additional data: $\alpha = 2$, robust approach, Equation \eqref{tensao_RTO}; $\beta_{T} = 2$, reliability-based approach, Equation \eqref{Problema_otm_5};
\item Non-probabilistic problem, where the uncertain magnitudes are bounded as: $f_{V} \in [6,14]$N and $f_{H} \in [-2,6]$N. The bounds are defined considering $\text{E}[f] \pm 2 \times \text{Std}[f]$ (since $\alpha = \beta_{T} = 2$ in the probabilistic problem).
\end{itemize}

The anti-optimization problems, related to the non-probabilistic formulation, are solved with the two step procedure described earlier. For each stress constraint: 1) the von Mises stress is evaluated at 9 points (combinations of $\mathbf{f}_{H}^{grid} = [-2,2,6]^{T}$N and $\mathbf{f}_{V}^{grid} = [6,10,14]^{T}$N); 2) from the point that presents the maximum von Mises stress, the modified steepest descent method is employed to achieve the solution of the anti-optimization problem.

Obtained results are post-processed through MCS, for $f_{H} \sim N (2,2)$N and $f_{V} \sim N (10,2)$N. Reliability indices, $\beta^{(k)} = -\Phi^{-1}\left(P_{f}^{(k)}\right)$, are evaluated at each point of stress computation $k$, where $P_{f}^{(k)}$ is obtained by dividing the number of failures at $k$ by the total number of realizations, $1 \times 10^{6}$. A maximum value of $\beta_{max} = -\Phi^{-1}\left(1 \times 10^{-6}\right) \cong 4.75$ is considered to illustrate the post-processed reliability indices.

\begin{figure*}[ht!]
\centering
\includegraphics[width=0.8\textwidth]{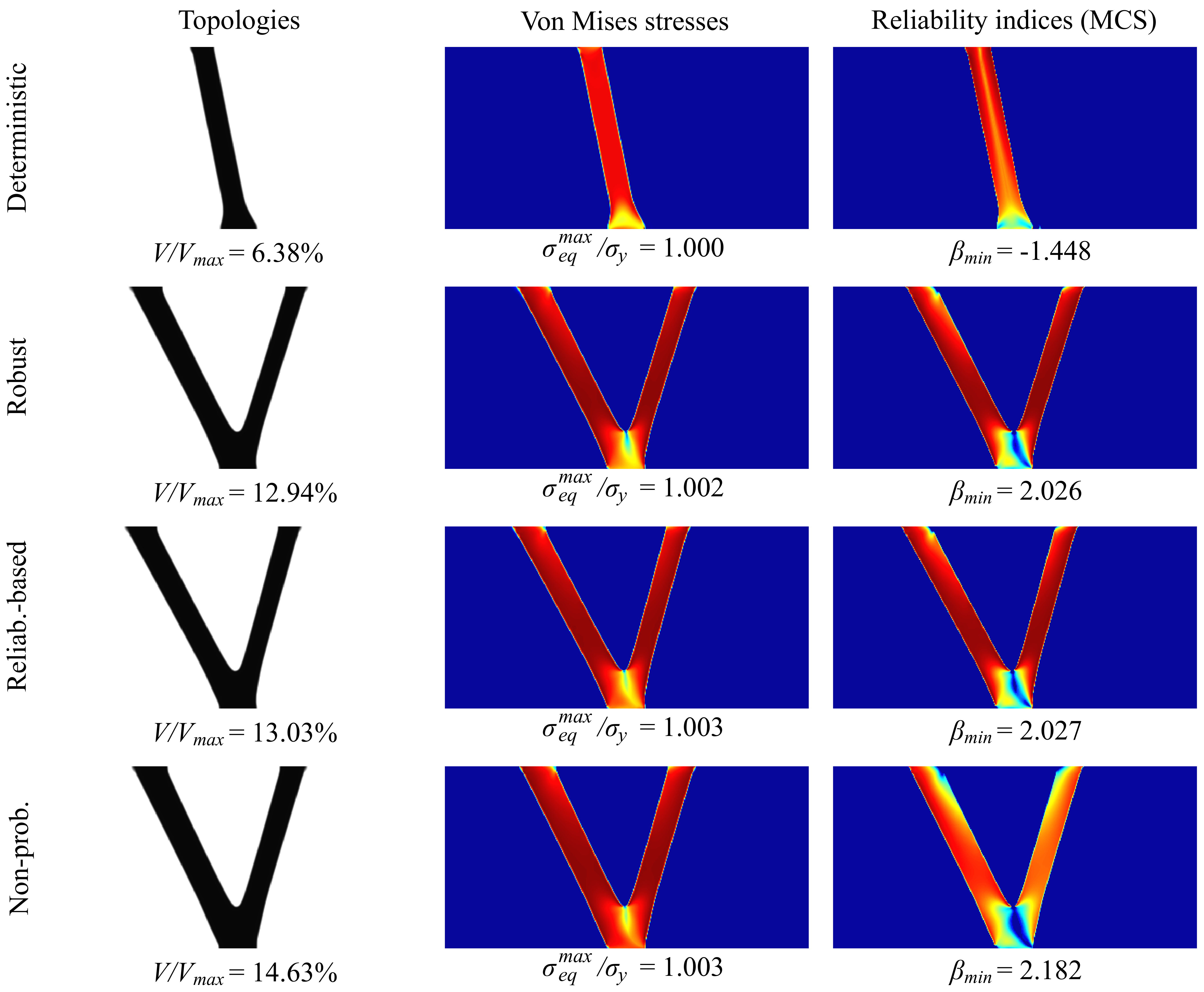}
\caption{Problem with rectangular design domain, results: topologies (left); von Mises equivalent stresses (middle); post-processed reliability indices (right). The stresses shown in the figure are the respective stress measures defined for each problem. Structural volumes, maximum stresses and minimum reliability indices are shown below each figure.}
\label{Resultados_retangular}
\end{figure*}

Figure \ref{Resultados_retangular} shows optimized topologies, von Mises equivalent stresses and post-processed reliability indices. The von Mises stresses illustrated in Figure \ref{Resultados_retangular} are the stress measures employed in each formulation: $\sigma_{eq}^{(k)}(\overline{\bm{\rho}})$ (deterministic); $\hat{\sigma}_{eq}^{(k)}(\overline{\bm{\rho}},\mathbf{Z})$ (robust); $\sigma_{eq}^{(k)}\left(\overline{\bm{\rho}},\left(\mathbf{Y}^{(k)}\right)^{*}\right)$ (reliability-based); and $\sigma_{eq}^{(k)}\left(\overline{\bm{\rho}},\left(\mathbf{W}^{(k)}\right)^{*}\right)$ (non-probabilistic).

By analyzing Figure \ref{Resultados_retangular}, one can observe that the topologies obtained with the non-deterministic approaches have two structural members, instead of only one member as the deterministic solution. However, although these topologies are the same (same number of holes), different shapes and structural volumes are obtained. Post-processed reliability indices demonstrate that the non-deterministic solutions are more reliable than the deterministic one, since their minimum reliability indices resulted next to the target reliability index defined for the reliability-based problem, i.e., $\beta_{min} \cong \beta_{T} = 2$, while the minimum reliability index for the deterministic solution resulted much smaller, as $\beta_{min} \cong -1.5$. Maximum probabilities of failure (related to the minimum reliability indices) are shown in Table \ref{pf_retangular}.

\begin{table}[ht!]
\caption{Structural volumes, minimum reliability indices, maximum probabilities of failure, and number of iterations until convergence. Rectangular problem.}
\label{pf_retangular}
\centering
\begin{tabular}{lcccc}
\hline
Problem & $V/V_{max}$ & $\beta_{min}$ & $P_{f}^{max}$ & Iterations \\
\hline
Deterministic & $6.38\%$  & $-1.448$ & $92.62\%$ & $942$ \\
Robust & $12.94\%$  & $2.026$ & $2.14\%$ & $1334$ \\
Reliability-based & $13.03\%$  & $2.027$ & $2.13\%$ & $1250$ \\
Non-probabilistic & $14.63\%$  & $2.182$ & $1.46\%$ & $1359$ \\
\hline
\end{tabular}
\end{table}

One can observe, in Figure \ref{Resultados_retangular}, in the stress graphs, that all structures are highly stressed, given the employed stress measures, and that the maximum stresses exceed the yield stress in less than $1\%$, indicating that all structures are truly optimized given the respective optimization problems; i.e., there is no room for improvements regarding the structural volume, since additional volume minimization would imply in higher stresses, and hence, unfeasible solutions.

Although the employed stress measures for robust and reliability-based approaches have different meanings, respective optimization solutions are almost the same. Post-processed reliability indices are next to 2 in both cases, indicating that the robust formulation, in this case, can be employed as an alternative approach to the reliability-based one. However, although very similar in this case, nothing can be said for other problems, specially when non-Gaussian variables are employed in representing the uncertain variables.

The non-probabilistic approach, on the other hand, provided a more conservative result when compared with the probabilistic ones, since its minimum reliability index, $\beta_{min} = 2.182$, resulted slightly larger than $\beta_{T} = 2$. This is justified, since the non-probabilistic approach ensures stress constraint feasibility for the whole interval defined by the combinations of the bounds of the uncertain loads. In the reliability-based approach, on the other hand, the most extreme points next to the combinations of the bounds of the uncertain variables are not considered, since their probabilities of occurrence are smaller than the admissible failure probability. Thus, in this example, one can verify there is no direct relation between probabilistic and non-probabilistic approaches, since very distinct results are obtained.

Table \ref{pf_retangular} shows the number of iterations for each case. It is verified that the non-deterministic approaches require more iterations until convergence. Moreover, the non-deterministic approaches require solution of additional equilibrium equations, one for each uncertain load \cite{Artigo3,Artigo4,Artigo5}, in order to obtain auxiliary displacement fields and then compute the respective stress measures. In addition, in both nested approaches, inner optimization problems are solved for each point of stress computation, at the beginning of each outer iteration. However, since the principle of superposition is employed, the computational cost for solving a given inner problem (inverse reliability analysis or anti-optimization problem) is negligible, since there is no need to solve the equilibrium equations for computing the von Mises stresses (and their derivatives with respect to the uncertain variables) at each inner iteration.

Although not considered in our implementations, parallel computing can drastically reduce the computational cost associated with the computation of MiPPs (reliability-based approach) or the worst-case scenario for the stress constraints (non-probabilistic approaches) at the beginning of each outer iteration. Since the inner problems do not depend on each other, these can be solved simultaneously, such that the use of parallel computing can be effective in reducing the total computational cost, as demonstrated by \cite{Parallel_computation}, in the design of MEMS (MicroElectroMechanical Systems) under unknown-but-bounded parameters.

\subsection{L-shaped problem under one uncertain load}

L-shaped design problems are often employed, in the literature, as benchmark problems, in order to test new algorithms and formulations for stress-based topology design \cite{ChauLe,Norato_level_set}. These problems are specially interesting, from a stress-based design point of view, since the L-shaped design domain has a sharp corner that leads to stress concentration, which should be properly avoided by the algorithm in order to ensure a rounded corner on the optimized topology.

The L-shaped design problem addressed in this subsection, Figure \ref{dominio_retangular_L1} (b), is discretized with $57600$ elements. Input data: filtering radius of $R=0.02$m and yield stress of $\sigma_{y} = 16$kPa. Two loads are applied, one horizontal and one vertical. The vertical load is deterministic, $f_{V} = 0.3$N, and the horizontal load has deterministic direction and uncertain magnitude, $f_{H} \sim N(0,0.015)$N. The applied loads are distributed over a length of $0.06$m to avoid stress concentration.

In order to properly compare the deterministic and non-deterministic approaches, three distinct problems are formulated, based on how the horizontal load is handled:
\begin{itemize}
\item Deterministic problem, with null horizontal load (mean value);
\item Probabilistic problem, considering $f_{H} \sim N(0,0.015)$N. Additional data: $\alpha = 2$, robust approach, Equation \eqref{tensao_RTO}; $\beta_{T} = 2$, reliability-based approach, Equation \eqref{Problema_otm_5};
\item Non-probabilistic problem, where the uncertain magnitude is bounded as: $f_{H} \in [-0.03,0.03]$N. The bounds are defined considering $\text{E}[f] \pm 2 \times \text{Std}[f]$ (since $\alpha = \beta_{T} = 2$ in the probabilistic problem).
\end{itemize}

The anti-optimization problems (related to the non-probabilistic approach) are solved with the two step procedure. For each stress constraint: 1) the von Mises stress is evaluated at 3 points ($\mathbf{f}_{H}^{grid} = [-0.03,0,0.03]^{T}$N); 2) from the point that presents the maximum von Mises stress, the modified steepest descent method is employed to achieve the solution of the anti-optimization problem.

\begin{figure*}[ht!]
\centering
\includegraphics[width=0.8\textwidth]{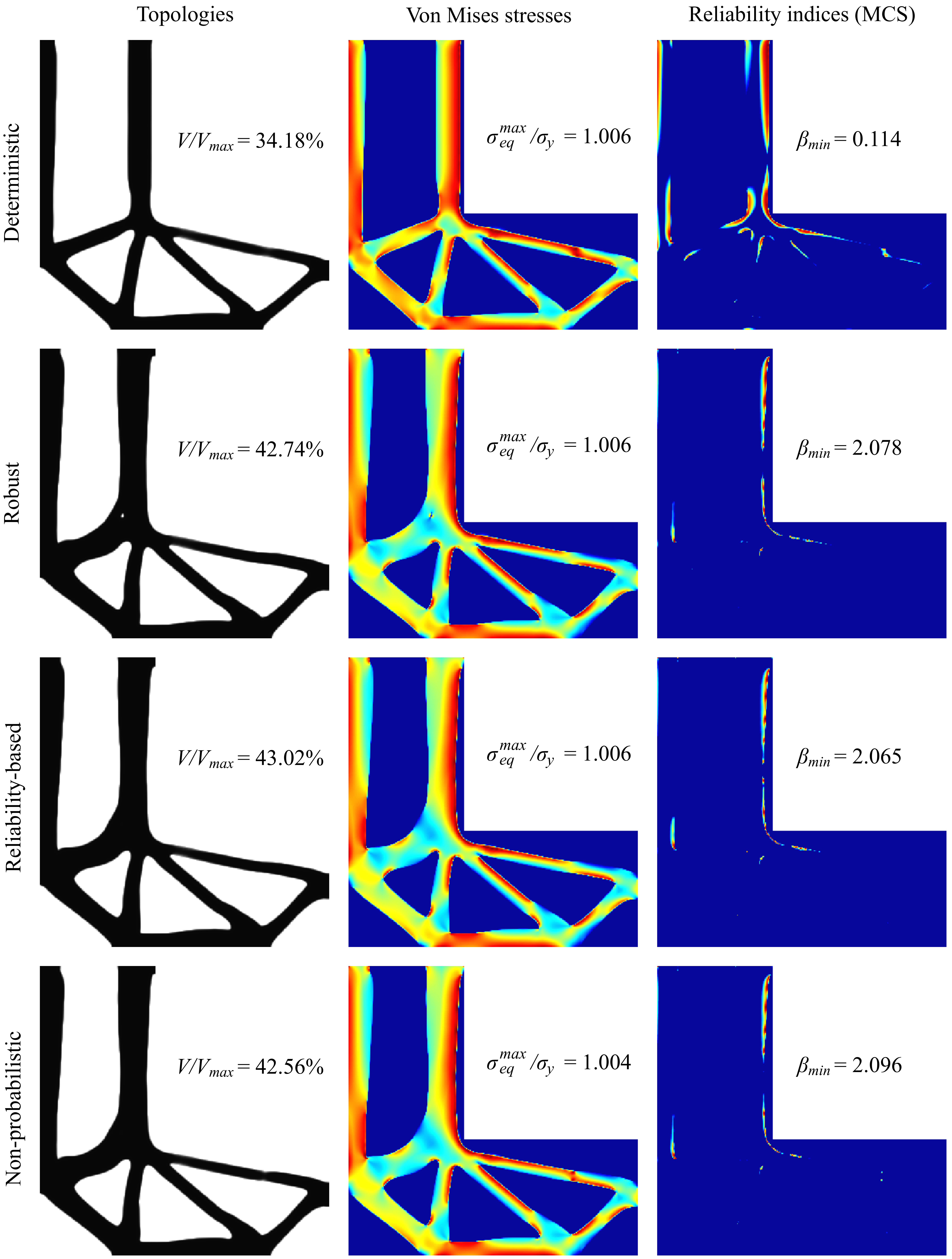}
\caption{L-shaped design problem, results: topologies (left); von Mises equivalent stresses (middle); post-processed reliability indices (right). The stresses shown in the figure are the respective stress measures defined for each problem.}
\label{Resultados_L1}
\end{figure*}

Figure \ref{Resultados_L1} shows the obtained results, including the post-processed reliability indices. One can verify that the obtained topologies have rounded corners, thus avoiding the sharp corner of the design domain; moreover, the stresses at these regions satisfy the stress failure criteria. The optimized topologies obtained as solution of deterministic, reliability-based and non-probabilistic problems are the same. The topology obtained as solution of the robust problem presents a small local difference, which consists in a small hole. The post-processed reliability indices demonstrate that the deterministic structure is extremely sensitive to variations in the horizontal load. On the other hand, it is shown that all the non-deterministic approaches provide reliable results, in the sense that their minimum post-processed reliability indices resulted next to the target reliability index, i.e., $\beta_{min} \cong \beta_{T}$.

Contrary to what happens in the previous problem (under two uncertain loads), the L-shaped problem subjected to one uncertain load only, when addressed with the non-probabilistic approach, does not provide a more conservative result when compared with the probabilistic approaches. The three structures obtained by employing the non-deterministic approaches have similar structural volumes and post-processed minimum reliability indices. This is justified, since now only one uncertain variable is being considered. In this case, the extreme points of the interval are also taken into account in the probabilistic approach, since the probability of occurrence of these points coincides with the probability associated with the target reliability index, $\beta_{T}$.

It is interesting to observe, in this case, that the post-processed reliability indices are critical (i.e., next to $\beta_{min}$) in a few points only, different from what happens in the problem of rectangular design domain, Figure \ref{Resultados_retangular}, where the post-processed reliability indices are critical in almost the whole structure (at least considering the probabilistic results). It should be noted that the problem with rectangular design domain is quite simple, from a stress-based design point of view, which facilitates the obtaining of highly stressed structures. The L-shaped design problem, on the other hand, is more challenging, requiring a more complex structure, with fewer points under the maximum (yield) stress, which implies in post-processing graphs with fewer critical points.

Table \ref{pf_L1} shows the maximum probability of failure, associated with the minimum post-processed reliability index through MCS, and the number of iterations until convergence, for each topology optimization problem solved in this subsection.

\begin{table}[ht!]
\caption{Structural volumes, minimum reliability indices, maximum probabilities of failure, and number of iterations until convergence. L-shaped design problem.}
\label{pf_L1}
\centering
\begin{tabular}{lcccc}
\hline
Problem & $V/V_{max}$ & $\beta_{min}$ & $P_{f}^{max}$ & Iterations \\
\hline
Deterministic & $34.18\%$  & $0.114$ & $45.47\%$ & $1675$ \\
Robust & $42.74\%$  & $2.078$ & $1.89\%$ & $1848$ \\
Reliability-based & $43.02\%$  & $2.065$ & $1.94\%$ & $1913$ \\
Non-probabilistic & $42.56\%$  & $2.096$ & $1.80\%$ & $2145$ \\
\hline
\end{tabular}
\end{table}

By comparing Tables \ref{pf_retangular} and \ref{pf_L1}, one can verify the L-shaped design problem requires a larger number of iterations until convergence. This is justified, since L-shaped design problems are more challenging, from a stress-based design point of view. Table \ref{pf_L1} shows the non-deterministic approaches are more costly than the deterministic one, in agreement with the results shown earlier in Table \ref{pf_retangular}.

\section{Concluding remarks}
\label{conclusoes}

This work presented a comparison of deterministic, robust, reliability-based and non-probabilistic approaches for stress-constrained topology optimization of continuum structures under uncertainty in applied loads. It is demonstrated that all approaches, deterministic and non-deterministic, can be formulated in a similar way, in which resulting optimization problems can be solved by the same method. However, each non-deterministic formulation requires an uncertainty propagation approach for handling the uncertainty in applied loads, such as: 1) the first-order perturbation method, to evaluate the expectation and standard deviation of von Mises stresses in the robust approach; 2) the PMA, to formulate the reliability-based approach; and 3) the anti-optimization approach, to obtain the worst-case scenario for the stress constraints in the non-probabilistic approach.

Two optimization problems were solved. Numerical results demonstrated that, while the deterministic solutions resulted extremely unreliable, all non-deterministic approaches resulted in robust and/or reliable structures, even though the formulations are completely different from each other considering a mathematical point of view.

Even though first-order approaches were employed to formulate the probabilistic problems, good agreement was obtained in the post-processing with MCS. Regarding the topology optimization problem under two applied loads with uncorrelated Gaussian magnitudes, it was shown that the robust and reliability-based approaches provide similar results, indicating the robust approach, in this case, can be employed as an alternative to the reliability-based approach. It was also shown that the non-probabilistic approach provides a more conservative result when compared with the other approaches, since it takes into account the combination of extremes of the intervals, ensuring stress constraint feasibility also for these cases, contrary to what happens in the probabilistic approaches, where the combinations of extremes are neglected due to their low probabilities of occurrence. When only one load of uncertain magnitude was considered, the three non-deterministic approaches provided similar results. This is justified, in this case, since there is no combination of extremes when only one uncertain variable is considered.

\appendix

\section*{Appendix: Density filtering with Heaviside step function}
\label{filtro_densidade}

In this paper, relative densities are not directly employed as design variables, i.e., they are not directly used during the optimization process. The use of relative densities as design variables leads to common problems in density-based topology optimization approaches, such as checkerboard-like areas and mesh dependent solutions \cite{BendsoeLivro}.

In order to alleviate these difficulties, density filtering with Heaviside step function is employed \cite{Sigmund-2007}. In this approach, relative density of element $e$ is computed as
\begin{linenomath}\begin{equation}
\overline{\rho}_{e} = 1 - \mathrm{e}^{-\delta \tilde{\rho}_{e}} + \tilde{\rho}_{e} \mathrm{e}^{-\delta},\label{projecao_Heaviside}
\end{equation}\end{linenomath}
where $\tilde{\rho}_{e}$ is the filtered relative density of element $e$, obtained from a linear projection
\begin{linenomath}\begin{equation}
\tilde{\rho}_{e} = \frac{\sum_{i \in \vartheta_{e}} w(\mathbf{x}_{i}) V_{i} \rho_{i}}{\sum_{i \in \vartheta_{e}} w(\mathbf{x}_{i}) V_{i}},\label{projecao_linear}
\end{equation}\end{linenomath}
over the design variables $\bm{\rho}$, in a circular neighborhood $\vartheta_{e}$, centred in element $e$, which contains all the elements whose center is within a radius $R$ specified by the designer.

A linear weighting function is employed and is defined as
\begin{linenomath}\begin{equation}
w(\mathbf{x}_{i}) = R - \| \mathbf{x}_{i} - \mathbf{x}_{e} \|,\label{decaimento_linear}
\end{equation}\end{linenomath}
where $\mathbf{x}_{i}$ contains the coordinates of the center of element $i$ and $\mathbf{x}_{e}$ contains the coordinates of the center of the neighborhood $\vartheta_{e}$.

For $\delta=0$, in Equation \eqref{projecao_Heaviside}, a linear behavior between physical and filtered relative densities is obtained, $\overline{\rho}_{e} = \tilde{\rho}_{e}$, whereas for $\delta \rightarrow \infty$, the Heaviside step function is obtained \citep{Guest_Heaviside}. As an additional benefit to avoid checkerboard patterns and ensuring mesh independent solutions, the employed density filter exploits use of a smoothed Heaviside step, Equation \eqref{projecao_Heaviside}, which is usually employed in order to achieve crisp black and white solutions by considering large values of $\delta$, reducing the blurred boundaries effect related to linear density filtering \citep{Sigmund-2007}.

\section*{Acknowledgements}

The authors acknowledge financial support of this research project by the agencies CNPq (National Council for Research and Development), grant number 306373/2016-5, FAPESP (S\~ao Paulo Research Foundation), grant number 2018/16701-1, and FAPESC, grant numbers 2017TR1747 and 2017TR784. This study was financed in part by the Coordination for the Improvement of Higher Education Personnel - Brazil (CAPES) - Finance Code 001.

\bibliographystyle{elsarticle-num}
\bibliography{bibliografia}





\end{document}